\documentclass[conference]{IEEEtran}
\IEEEoverridecommandlockouts
\usepackage{cite}
\usepackage{amsmath,amssymb,amsfonts}
\usepackage{algorithm}
\usepackage{algorithmic}
\usepackage{graphicx}
\usepackage{textcomp}
\usepackage{xcolor}
\usepackage{ulem}
\def\BibTeX{{\rm B\kern-.05em{\sc i\kern-.025em b}\kern-.08em
    T\kern-.1667em\lower.7ex\hbox{E}\kern-.125emX}}
\begin{document}

\newcommand{\ut}[1] {\textbf{UT: #1}}

\title{Measuring scheduling efficiency of RNNs for NLP applications \\
}


\author{\IEEEauthorblockN{Urmish Thakker,
Ganesh Dasika, Jesse Beu and
Matthew Mattina}
\IEEEauthorblockA{Arm ML Research Lab}
}


\maketitle

\begin{abstract}
Recurrent neural networks (RNNs) have shown state of the art results for speech recognition, natural language processing, image captioning, and video summarizing applications. Many of these applications run on low-power platforms, so their energy efficiency is extremely important. We observed that cache-oblivious RNN scheduling during inference typically results in 30-50x more data transferred on and off the CPU than the application's working set size. This can potentially impact its energy efficiency. This paper presents a new metric called Data Reuse Efficiency to gauge the RNN scheduling efficiency of a platform and shows the factors that influence the DRE value. Additionally, this paper discusses an optimization to improve reuse in RNNs and highlights the positive impact of this optimization on the total amount of memory read from or written to the memory controller (and, hence, the DRE value) during the execution of an RNN application for a mobile SoC. 
\end{abstract}


\section{Recurrent Neural Networks (RNNs)}
\label{rnn}
Recurrent neural networks are a type of deep neural network (DNN) that make use of sequential information. They are used in tasks where the ordering of the input sequence is important (e.g., time-based data, natural language processing). The fundamental component of a RNN is a cell. The cells have weights and an internal state. The state is updated by applying the same computation and weights to every element of a sequence, in sequence order, over multiple time steps. This state is called the hidden vector and acts as a "memory". Multiple cells can be stacked on top of each other to form a multilayer recurrent neural network. Most popular RNN cell types are long short-term memories (LSTMs)\cite{b3} and gated recurrent units (GRUs)\cite{b4}.

\begin{figure}[th]
  \includegraphics[width=\linewidth]{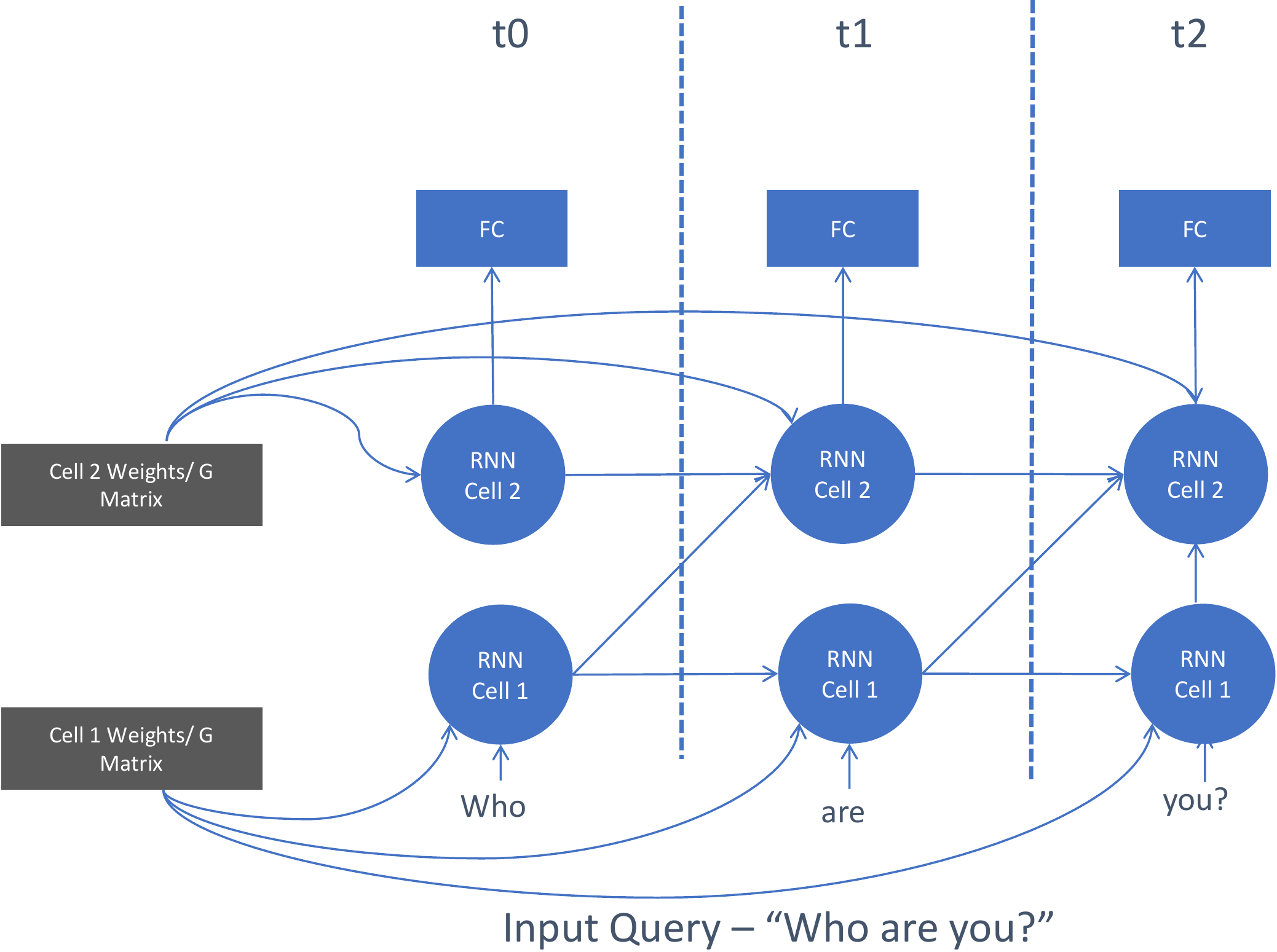}
  \caption{2 layer RNN network}
  \label{fig:fig1}
\end{figure}

\begin{figure}[t]
  \includegraphics[width=\linewidth]{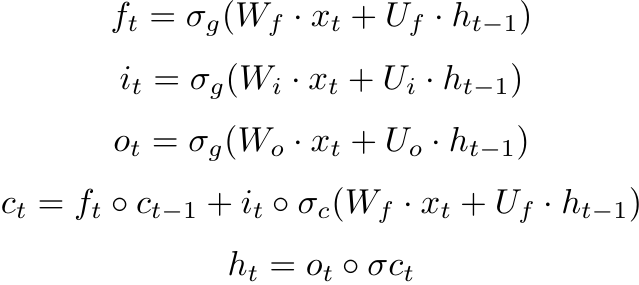}
  \caption{Equations executed by a LSTM layer}
  \label{fig:fig2}
\end{figure}

Figure~\ref{fig:fig1} represents a 2-layer RNN network followed by a fully connected softmax layer. The input to the network is the query, ``Who are you?'' spread over three time steps. The weights of RNN Cell 1 and 2 are denoted by ``Cell Weights 1'' and ``Cell Weights 2''. These cell weights do not change across each time step. Figure~\ref{fig:fig2} shows the equations executed by a LSTM layer. Here $x_t$ denotes the input at time step $t$, $h_{t-1}$ denotes the memory element from previous time step and $h_t$ denotes the memory element after the end of the computation during the current time step. The cell weights in Figure~\ref{fig:fig1} are the concatenation of $W$ and $U$ matrices in Figure~\ref{fig:fig2}. 

\section{Current RNN Scheduling}
\label{intro:nonstream}

To efficiently execute the operations in Figure~\ref{fig:fig2}, these weight matrices are concatenated together to create a larger matrix (referred to as $G$ in this paper) which is then multiplied by the concatenation of the input vector and the hidden vector.

\begin{algorithm}[t]
  \caption{\textbf{Scheduling algorithm for non-streaming applications:} Each input $x\textsubscript{0:T}$ is concatenated with the hidden vector from the previous step, $h$ and multiplied with G. The result passes through a non-linearity layer, after which it is split into 4 sub-vectors. These sub-vectors are linearly combined to generate the hidden vector for the current time step.}\label{alg:nonstream}
  \begin{algorithmic}[1]
    \STATE {Input - $G,T,n,x\textsubscript{0:T}$}
    \STATE $r\gets 1$
    \STATE $c\gets 0$
    \STATE $h\gets 0$
    \STATE $I\gets 0$
    \FOR{\texttt{$r < T$}} 
      \STATE $I = concat(x\textsubscript{r},h)$ 
      \STATE $Y= G*I$
      \STATE $Y'= sig(Y'[0:3*n,:])$
      \STATE $f = Y'[0:n]$
      \STATE $i = Y'[n:2*n]$
      \STATE $o = Y'[2*n:3*n]$
      \STATE $c' = tanh(Y[3*n:4*n,0])$
      \STATE $c=f\circ c + i\circ c'$
      \STATE $h=o \circ c$
      \STATE $r=r+1$
    \ENDFOR 
    \STATE \textbf{return} $h$
  \end{algorithmic}
\end{algorithm}

\[
G=
  \begin{bmatrix}
      W\textsubscript{f} & W\textsubscript{i} & W\textsubscript{o} & W\textsubscript{c} \\
      U\textsubscript{f} & U\textsubscript{i} & U\textsubscript{o} & U\textsubscript{c} 
  \end{bmatrix}
\]

\[
Concatenated Input Vector (I)=
  \begin{bmatrix}
      x\textsubscript{t} \\
      h\textsubscript{t-1} 
  \end{bmatrix}
\]

G is a 2-row concatenation of the $W$ and $U$ matrices, where each row has 4 $W$ or $U$ matrices. Thus, if a single $W$ or $U$ matrix is of size [\textit{n,n}], then the concatenated matrix, G, has the size [\textit{2*n,4*n}].

Following the equations in Figure~\ref{fig:fig2}, $G$ is multiplied by $I$. The resultant output vector is divided into 4 vectors to execute the rest of the operations. 

The current schedule for inference of a multilayered LSTM network is:
\begin{enumerate}
	\item{An input $I$ is broken down into multiple vectors: $x_1$, $x_2$, $x_3$, \ldots, $x_T$.}
	\item{The computations are scheduled as described in Algorithm~\ref{alg:nonstream}}
	\item{Step 2 is repeated for each layer in the multilayered RNN.}
\end{enumerate}

The above schedule is not restricted to LSTMs only and is applicable to any RNN Cell network. 

\section{Issues with current RNN scheduling algorithms}
\label{sec:issues}

To identify the issue with the current scheduling scheme, we run experiments on a desktop Intel Haswell Platform with a 25 MB last-level cache (LLC). We use benchmarks written in TensorFlow 1.4 compiled with the Intel MKL library. An optimized dataflow graph for deployment is created using the freeze\_graph tool in Tensorflow and the resultant graph is executed to process the input queries.

Table \ref{table:tab1} lists the configuration of the LSTM benchmarks evaluated. All the parameters in Table~\ref{table:tab1} are combined to create a total of 320 benchmarks. The basic architecture for each benchmark is similar to that shown in Figure~\ref{fig:fig1} -- multiple RNN layers followed by a softmax layer.  The benchmarks cover small and large vocabularies. Vocabulary is the number of words/tokens used in a NLP application. Each word/token is represented via a vector and the matrix of vectors of all the words/tokens used by the application is called an embedding. For NLP applications, embeddings are present in the first and the last layers of the network.  

\begin{table*}[]
\begin{center}
\begin{small}
\begin{tabular}{|l|p{8cm}|p{5cm}|}
\hline
\multicolumn{1}{|c|}{Config. Name} & \multicolumn{1}{c|}{Description} & \multicolumn{1}{c|}{Value} \\ \hline
	RNN Size & Size of the hidden vector & 64,128,256,512,1024 \\ \hline
Number of Layers & Number of stacked RNN Cells on top of each other & 1,2,3,4,5,6,7,8 \\ \hline
RNN Cell Type & Type of RNN cell & LSTM, GRU \\ \hline
Length of Input & Number of time steps over which input is fed to the RNN network & 1,10,50,100 \\ \hline
	Vocabulary Size &  Number of classes in the dictionary & Small - 60, Large - 10,000 \\ \hline
	Batch Size & Number of inferences simultaneously processed & 1  \\ \hline
\end{tabular}
\end{small}
\end{center}
\caption{Benchmark Parameters (Total of 320 benchmarks)}
\label{table:tab1}
\end{table*}

\subsection{Data Reuse Efficiency (DRE)}
This paper introduces a new metric to measure the efficiency of scheduling RNN operations in a modern deep learning framework. The metric measures how much more data than the working set of the RNN application was used during the execution of the RNN application. Thus, efficiency in this context is an indication of the usage of memory bandwidth. This metric requires measuring two quantities - bandwidth usage of the application and application working set size. The bandwidth usage of an application can be measured by using the performance counters in the memory controller of the platform. When measuring the bandwidth usage, each benchmark is run 100 times. The average amount of data read from and written to a memory controller ($AvgRW$) across all runs is measured. Next, the working set (weights + word embeddings + intermediate values) of the RNN network is calculated from the benchmark configuration parameters. The inefficiency in scheduling is measured by calculating the ratio, 

\begin{equation}
DataReuseEfficiency(DRE) = AvgRW/WorkingSet
\end{equation}

The lower the value of $DRE$, the better a scheduling algorithm. An ideal system will have a $DRE$ of 1 i.e., it is able to cache all the weights until they are no longer required.

\subsection{Results}

\begin{figure}[t]
  \includegraphics[width=\linewidth]{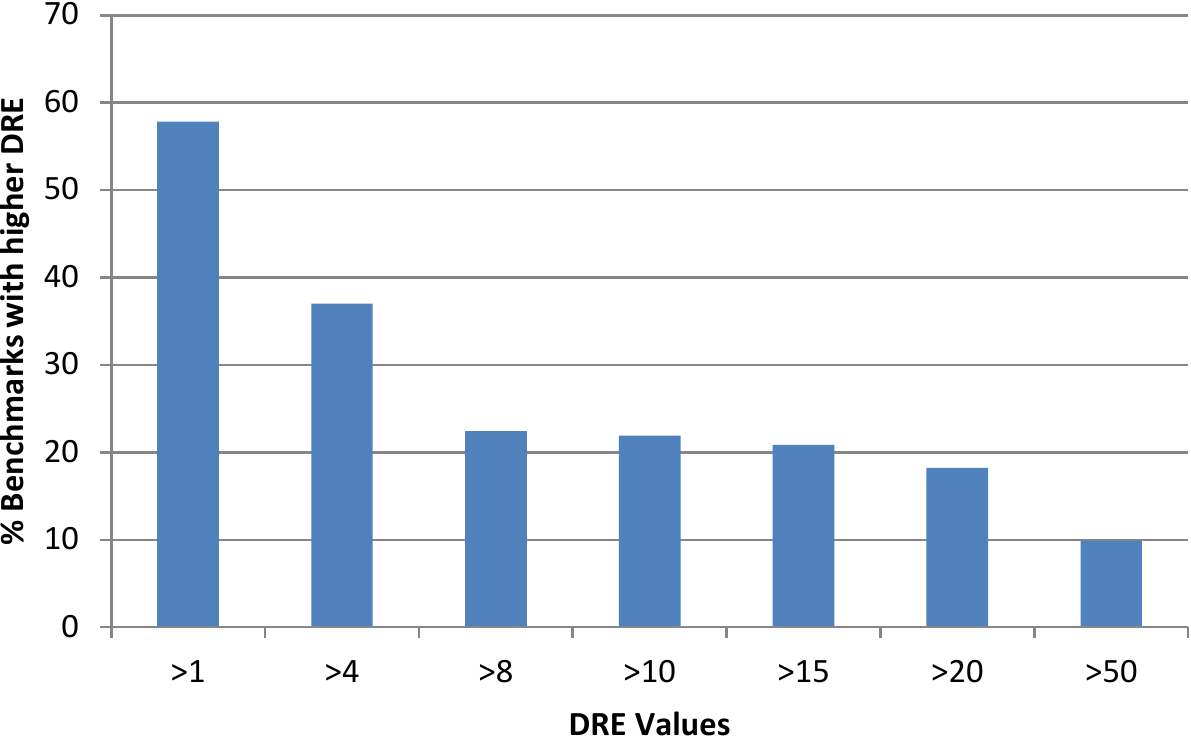}
	\caption{Distribution of DRE values for all benchmarks with small vocabulary (60 words).} 
  \label{fig:smDRE}
\end{figure}

\begin{figure}[t]
  \includegraphics[width=\linewidth]{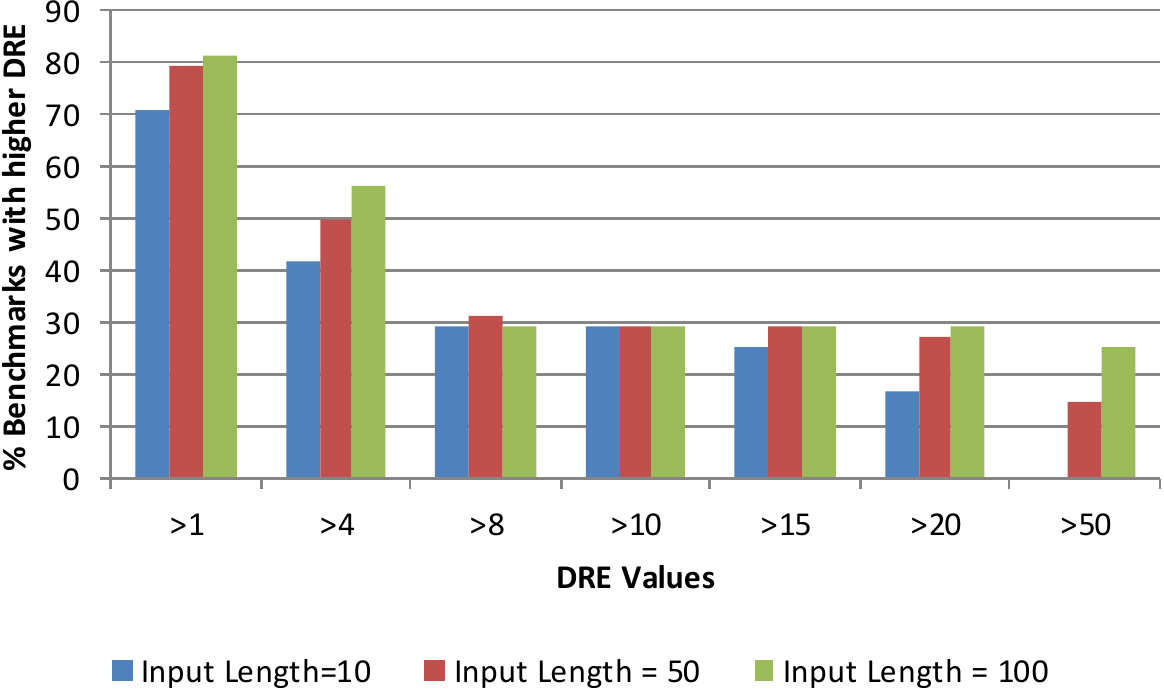}
	\caption{Distribution of DRE values for all benchmarks with small vocabulary (60 words). The distribution divided based on input length.} 
  \label{fig:smildre}
\end{figure}

{\bf LSTM (with small vocabulary):} For small vocabulary, the total size of output embedding does not exceed 0.5 MB. Figure~\ref{fig:smDRE} shows the \% of benchmarks exceeding a certain $DRE$ value. The benchmarks that have a $DRE$ value $> 1$, have a working set size close to or greater than the size of LLC. 41\% of the benchmark's working set size is $> 25$ MB. Accounting for overheads (TensorFlow, python, OS, perf, and shell scripts used to run experiments and collect results), the benchmarks whose working set size is close to 25 MB will also not fit in the cache. These benchmarks will compete with overheads for space in the cache. If we assume that the overheads take up 3 MB of space in the cache, a working set size of 22 MB (16\% of the benchmarks) or more will not fit in the cache completely. Thus a total of 57\% of benchmarks have a $DRE$ value $> 1$. Other benchmarks, along with the overheads, completely fit in the cache and do not see any read write traffic at the memory controller. Since $DRE$ captures the $AvgRW$ traffic across 100 runs, the value of $AvgRW$ will be less than the working set of these benchmark. 

There is also a strong correlation between the length of input and the $DRE$ value. Figure~\ref{fig:smildre} breaks down the values in Figure~\ref{fig:smDRE} based on input length. It shows a bar chart of \% of benchmarks with input of length 10 to 100 and satisfying a $DRE$ criterion. The first blue bar of Figure ~\ref{fig:smildre} implies that 70.8\% of benchmarks that have an input length of 10, have a $DRE$ value $> 1$ while 22\% of benchmarks that have an input length of 100, have a $DRE$ value $> 50$.

\begin{figure}
  \includegraphics[width=\linewidth]{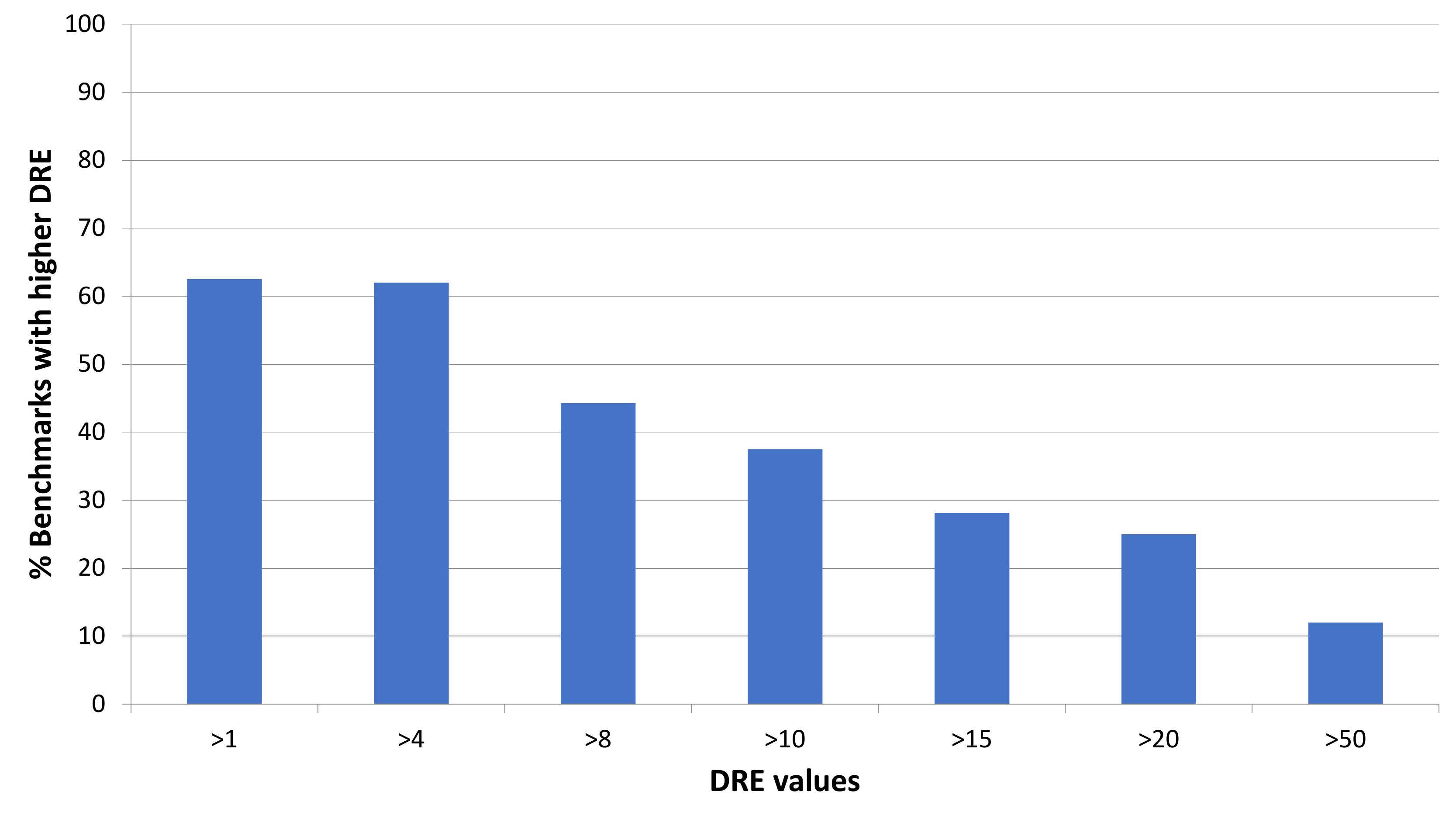}
	\caption{Distribution of DRE values for all benchmarks with large vocabulary  (10,000 words).} 
  \label{fig:largeDRE}
\end{figure}

{\bf Impact of a larger vocabulary:} There is also a correlation between $DRE$ value and the vocabulary size. When the vocabulary size is 10, 000 words, the word embedding could be anywhere between 4.78 MB for hidden unit vector of size 64 and 72 MB for hidden unit vector of size 1024. Figure~\ref{fig:largeDRE} shows that with a larger vocabulary, the number of benchmarks that have a $DRE$ value $> 1$ and $DRE$ value $> 4$ increases to 62\%.

{\bf Results for GRU:} The results for GRU cell based networks are very similar. For brevity, we do not discuss them in the paper.

\subsection{Key Observations}
\begin{itemize}
\item The current scheduling algorithm does not focus on data reuse in RNNs during inference -- for some benchmarks, 50x more data is read than the working set of the benchmark.
\item For smaller caches (as seen in mobile devices), the problem of high $DRE$ should be even more prominent
\item The $DRE$ value is proportional to the number of layers, size of the hidden vector, vocabulary size and the length of the input. It increases as each of these values increases.
\end{itemize}

\section{Improving the bandwidth efficiency of RNN Cells}
\label{sec:breakingg}
{\bf Breaking the $G$ Matrix:} Reuse in RNNs is not obvious when we schedule them as discussed in Section~\ref{intro:nonstream}. However, across all the time steps, the matrix $G$ does not change. Matrix $G$ is composed of 2 sets of matrices. $W_{f/i/o/c}$ gets multiplied with inputs $x_1$, $x_2$, $x_3$, \ldots, $x_T$ while $U_{f/i/o/c}$ gets multiplied with the hidden vectors generated at each time step. 

Generally for NLP applications, the inputs are available before hand. Even if the input is streaming, buffering of inputs can create a scenario where inputs across some of the time steps are available beforehand. This provides an opportunity to implement a bandwidth efficient scheduling scheme by making three changes to the algorithm discussed in section~\ref{intro:nonstream}:
\begin{itemize}
\item $G$ should be broken down into two sets of matrices - $G1$ and $G2$.
\[
G1=
  \begin{bmatrix}
      W\textsubscript{f} & W\textsubscript{i} & W\textsubscript{o} & W\textsubscript{c} \\
  \end{bmatrix}
\]
\[
G2=
  \begin{bmatrix}
      U\textsubscript{f} & U\textsubscript{i} & U\textsubscript{o} & U\textsubscript{c} 
  \end{bmatrix}
\]
\item Next, the inputs should be concatenated across all time steps into a single matrix.  

\[
I'=
  \begin{bmatrix}
      x\textsubscript{1} & x\textsubscript{2} & x\textsubscript{3} & . & . & . & x\textsubscript{T} 
  \end{bmatrix}
\]
\item Finally, the computations should be executed as described in Algorithm~\ref{best}.
\end{itemize}

\begin{algorithm}
  \caption{\textbf{Bandwidth efficient LSTM computation:} The input elements $x\textsubscript{0:T}$ are all concatenated into a single matrix $I'$. This matrix gets multiplied by G1 and the result is stored in $X'$. For each time step, G2 is multiplied with the hidden vector from the previous time step and the result is added with the column representing the output of G1*I' for that time step. This vector passes through a non-linearity layer and gets split into multiple sub-vectors which are linearly combined to generate the hidden vector for the current time step.}\label{best}
  \begin{algorithmic}[1]
    \STATE {Input - $G1,G2,T,I',n$}
    \STATE $r\gets 0$
    \STATE $c\gets 0$
    \STATE $h\gets 0$
    \STATE $X'\gets G1*I'$ \label{alg:line:blah}
    \FOR{\texttt{$r < T$}} \label{alg:for:start}
      \STATE $Y'= G2*h$
      \STATE $Y'= X'[:,r]+Y'$ \label{alg:line:blah2}
      \STATE $Y'= sig(Y'[0:3*n,:])$
      \STATE $f = Y'[0:n]$
      \STATE $i = Y'[n:2*n]$
      \STATE $o = Y'[2*n:3*n]$
      \STATE $c' = tanh(Y[3*n:4*n,0])$
      \STATE $c=f\circ c + i\circ c'$
      \STATE $h=o \circ c$
      \STATE $r=r + 1$
    \ENDFOR \label{for:end}
    \STATE \textbf{return} $h$
  \end{algorithmic}
\end{algorithm}

Ideally, a scheduler should first schedule computation in line 5. Now, instead of recalculating $W_f$ $\cdot$ $x_t$, $W_i$ $\cdot$ $x_t$, $W_o$ $\cdot$ $x_t$ and $W_c$ $\cdot$ $x_t$ at every time step, the scheduling algorithm could read the $t^{th}$ column of $X'$ and use that as the input to subsequent computations (line 8). By doing the computation this way, we read $G1$ once. Next, the computations for line 7-15 are scheduled. These computations need to be scheduled for each vector in the input sequence. Thus, $G2$ still needs to be read at every time step because of the sequential dependency on the hidden vector (line 7). 

Assuming a input vector and hidden vector size of 512 each, the total size of the $G$ matrix is 8 MB (4 W and 4 U matrices, each of size 1 MB). If the size of the cache is smaller than 8 MB, $G$ matrix will not fit in the cache. For a system using the scheduling scheme described in Section~\ref{intro:nonstream}, the $G$ matrix will be reread from the memory at every time step. If the input length is 100, the total amount of data read from the memory is 800 MB. By using the new schedule, we can reduce the amount of data read from the memory to 404 MB instead of 800 MB, thus improving the $DRE$ value.

A similar scheduling scheme is discussed in \cite{b8}. However, this work differs from \cite{b8} in two ways. Firstly, they discuss this scheduling scheme from the perspective of a GPU. Secondly, to enable enough parallelism, they do two computations of input time steps simultaneously instead of all time steps in the above implementation (line 5, algorithm ~\ref{best}). Lastly, they do not look at the impact of this scheduling scheme on the memory bandwidth consumption.

\section{Evaluating the impact of the optimized scheduler}
We developed an analytical model to gauge the impact of the optimized scheduler on the memory system. It takes in the network configuration to create a data flow graph based on a scheduling algorithm. The data flow graph generates the memory requests which are fed to a least recently used (LRU) cache of size 12 MB. The cache size is chosen to mimic a modern day mobile SoC. Cache misses, along with writeback, are used to measure the read and write traffic to and from the memory. The matrix vector multiplication operations generate requests assuming cache blocking techniques to maximize reuse in the cache. 

\subsection{Benchmarks Evaluated}
The 4 applications modelled are language translation (Google's Neural Machine Translation (GNMT) \cite{b6}), speech recognition (DeepSpeech1 \cite{b5}), language modelling (LM) \cite{b8} and named entity recognition (byteNER \cite{b7}). GNMT has 8 encoder-decoder RNN layers along with an attention module and a vocabulary of size 80,000. DeepSpeech1 has 5 RNN layers and uses a vocabulary size of 28 words. LM consists of 2 LSTM layers and uses a vocabulary of size 80,000. byteNER has 4 RNN layers and uses a vocabulary of size 4. 

The benchmarks are evaluated using the following scheduling algorithms:
\begin{itemize}
\item {\bf Schedule A} - This schedule is described in Section~\ref{intro:nonstream} of this paper and is the one used in TensorFlow compiled with MKL. Computations are scheduled one layer at a time. Each layer processes inputs across all the time steps before moving to the next layer.
\item {\bf Schedule A+} - This is the schedule discussed in Section~\ref{sec:breakingg} where the $G$ matrix is broken into two matrices $G1$ and $G2$. 
\end{itemize}

\subsection{Results}
\label{results:nonstream}
\begin{figure}
  \includegraphics[width=\linewidth]{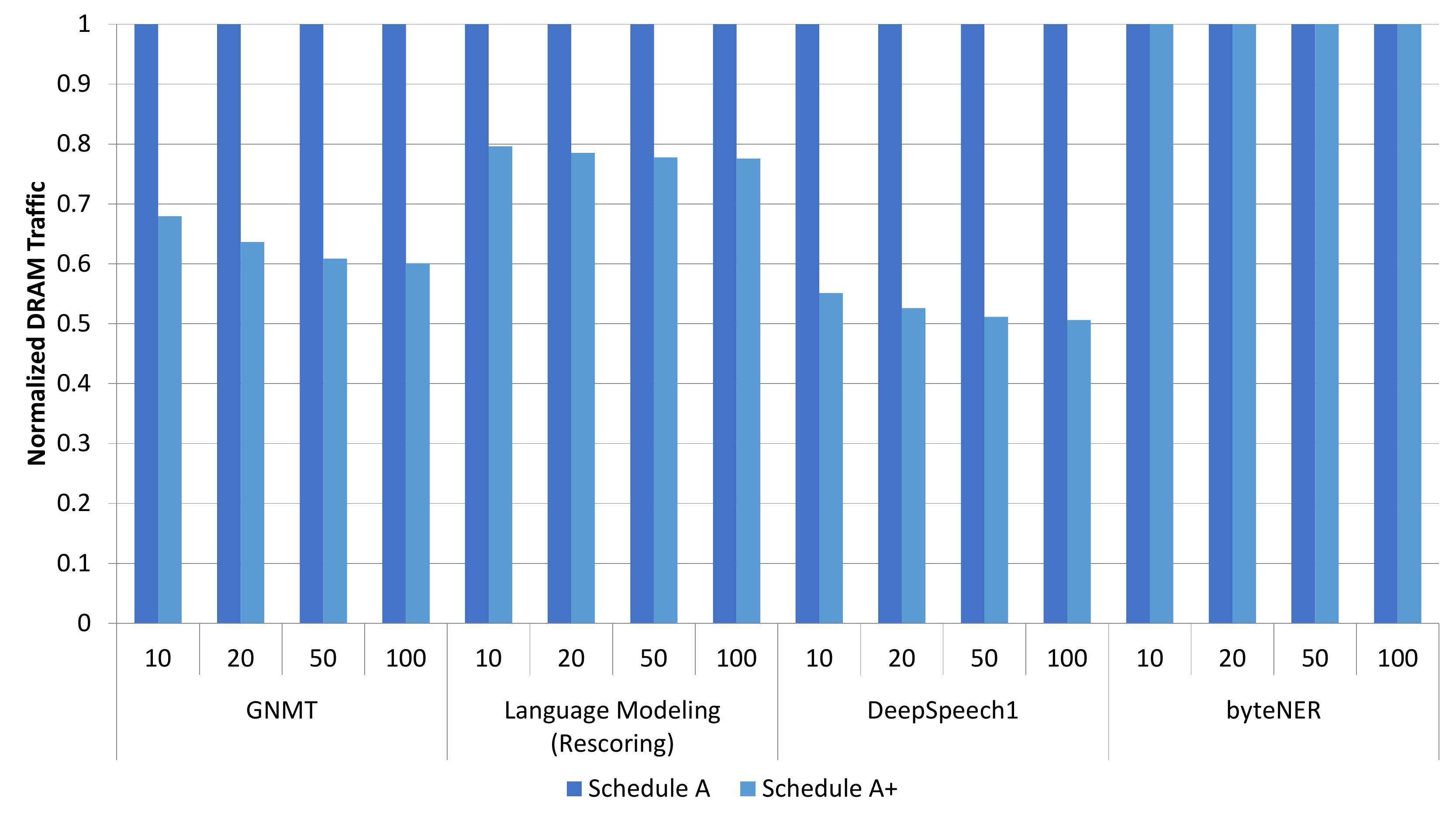}
  \caption{Bandwidth savings after breaking the G matrix for different scheduling algorithms and input lengths}
  \label{fig:nonstreambw}
\end{figure}

{\bf Impact on Bandwidth:} Figure~\ref{fig:nonstreambw} shows the impact of the optimized schedule, measuring the memory traffic to and from the memory. The values 10, 20, 50, and 100 indicate the number of time steps across which the input is fed to the network. The values have been normalized to schedule A's traffic for each benchmark. 

For LM, traffic to the memory reduces by $1.25 \times$ for smaller input lengths and $1.29\times$ for longer input lengths. LM uses cells of size 2,048 for one layer and 8,192 for the other. The $G1$ and $G2$ matrices are equally sized for the first layer. This gets 2x improvement for those layers. However, for the second layer, $G2$ is larger than $G1$ as each hidden-hidden weight matrix is of size 8192x8192 while the input weight matrix is of size 2048x8192. The optimization discussed in Section~\ref{sec:breakingg} reduces the cost of reading $G1$ multiple times. Since $G1$ is 4 times smaller than $G2$, A+ is not as effective for layer 2 as for layer 1. Thus, the overall benefit of A+ does not reach the expected factor of 2. Additionally, the $G$ matrix is 96 MB for the first layer and 960 MB for the second layer. These values will not fit in the cache. Thus, schedule A will not see any reuse.

GNMT is composed of encoder and decoder RNN layers, attention layers and word embedding layers. The $G$ matrix splitting does not apply to attention layers. The decoder layers need the output of the previous time step as an input to the next time step. As a result, the $G$ matrix splitting optimization cannot be applied to the decoder layers. Thus, A+ is only applied to the encoder layers, resulting in an improvement over schedule A by a factor of 1.45 to 1.65. 

ByteNER is a smaller network of size 9.4 MB. As a result, the network will fit in the cache and there will be significant reuse of data structures within the cache across multiple time steps for all the schedules. Thus, both schedules perform equally well. 

\section{Conclusion}
This paper introduces $DRE$, a new metric to measure the efficiency in scheduling RNN applications on CPUs. Using this metric, we have uncovered that typical RNN applications communicate with memory $30$-$50\times$ their working set size, due to inefficient data organization and scheduling. To counter this, this paper also introduced a new optimization to significantly improve the $DRE$ value and, consequently, improve the memory utilization efficiency.


\begin{thebibliography}{00}
\bibitem{b1} G. Eason, B. Noble, and I. N. Sneddon, ``On certain integrals of Lipschitz-Hankel type involving products of Bessel functions,'' Phil. Trans. Roy. Soc. London, vol. A247, pp. 529--551, April 1955.
\bibitem{b3} S Hochreiter, J Schmidhuber, ``Long Short-Term Memory," Neural Computation, 9(8):1735–1780, 1997. URL https://doi.org/10.1162/neco.1997.9.8.1735.
\bibitem{b4} K Cho, B Merrienboer, C Gehre, F Bougares, H Schwenk, H Schwenk, Y Bengio ``Learning Phrase Representations using {RNN} Encoder-Decoder for Statistical Machine Translation" CoRR, abs/1406.1078, 2014. URL http://arxiv.org/abs/1406.1078.
\bibitem{b5} A Hannun, C Case, J Casper, B Catanzaro, G Diamos, E Elsen, R Pregner, S Satheesh, S Sengupta, A Coates, A Ng, ``Deep Speech: Scaling up end-to-end speech recognition" CoRR, abs/1412.5567, 2014. URL http://arxiv.org/abs/1412.5567.
\bibitem{b6} Y Wu, M Schuster, Z Chen, Q Le, M Norouzi, W Macherey, J Dean et al, ``Google's Neural Machine Translation System: Bridging the Gap between Human and Machine Translation"
\bibitem{b7} Dan G, Cliff B, Oriol V, Amarnag S ``Multilingual Language Processing From Bytes" CoRR, abs/1609.08144, 2016. URL http://arxiv.org/abs/1609.08144.
\bibitem{b8} R Jozefowicz, O Vinyals, M Schuster, N Shazeer, Y Wu ``Exploring the Limits of Language Modeling" CoRR, abs/1602.02410, 2016. URL http://arxiv.org/abs/1602.02410.
\end{thebibliography}
\end{document}